\documentclass[aps,prl,reprint,balancelastpage,nofootinbib,preprintnumbers,superscriptaddress]{revtex4-1}

\usepackage{slashed}
\usepackage{bm} 
\usepackage{latexsym,amssymb,amsmath,float,url,mathrsfs}
\usepackage{latexsym}
\usepackage{graphicx}
\usepackage{epstopdf}
\usepackage{amsmath}
\usepackage{comment}
\usepackage{subfigure}
\usepackage{natbib}
\usepackage{hyperref}
\usepackage{pifont}
\usepackage{blindtext}
\usepackage{adjustbox}
\usepackage{multirow}
\usepackage{tabularx}
\usepackage[table]{xcolor}
\usepackage{orcidlink}
\usepackage{tikz}
\usetikzlibrary{tikzmark}
\usepackage{fontawesome}

\usepackage{blkarray}
\usepackage{graphicx}
\usepackage{amsfonts}
\usepackage{soul}
\usepackage{amssymb}
\usepackage{amsmath}
\usepackage{cancel}
\usepackage{tcolorbox}
\usepackage{tikz-feynman}
\usepackage{tcolorbox}

\usepackage{graphics,appendix,afterpage,makecell} 

\definecolor{oucrimsonred}{rgb}{0.6, 0.0, 0.0}
\definecolor{persianblue}{rgb}{0.11, 0.22, 0.73}
\definecolor{forestgreen}{rgb}{0.13,0.35,0.13}
\definecolor{lightgray}{rgb}{0.83, 0.83, 0.83}
 \hypersetup{colorlinks, citecolor=oucrimsonred, linkcolor=black, urlcolor=oucrimsonred}
\definecolor{cornellred}{rgb}{0.7, 0.11, 0.11}
\definecolor{navyblue}{rgb}{0.0, 0.0, 0.5}
\definecolor{amethyst}{rgb}{0.6, 0.4, 0.8}
\definecolor{yellow}{rgb}{1.0, 1.0, 0.0}
\definecolor{firebrick}{rgb}{0.7, 0.13, 0.13}
\definecolor{tangerineyellow}{rgb}{1.0, 0.8, 0.0}
\definecolor{deepfuchsia}{rgb}{0.76, 0.33, 0.76}
\definecolor{amber}{rgb}{1.0, 0.75, 0.0}
\definecolor{VioletRed4}{rgb}{0.55, 0.13, .32}
\definecolor{indiagreen}{rgb}{0.07, 0.53, 0.03}
\definecolor{VioletRed4}{rgb}{0.55, 0.13, .32}
\newcommand{\be}{\begin{equation}}
\newcommand{\ee}{\end{equation}}
\newcommand{\bea}{\begin{equation} \begin{aligned}}
\newcommand{\eea}{\end{aligned} \end{equation}}

\definecolor{oucrimsonred}{rgb}{0.6, 0.0, 0.0}
\newcommand\vertarrowbox[3][6ex]{%
  \begin{array}[t]{@{}c@{}} #2 \\
  \left\uparrow\vcenter{\hrule height #1}\right.\kern-blldelimiterspace\\
  \makebox[0pt]{\scriptsize#3}
  \end{array}%
}

\definecolor{verdechiaro}{rgb}{0.6,1,0.6}
\definecolor{giallochiaro}{rgb}{1,1,0.6}
\definecolor{bluscuro}{rgb}{0.15, 0.2, 0.9}
\definecolor{verdes}{rgb}{0.1, 0.5, 0.1}%
\definecolor{tangerineyellow}{rgb}{1.0, 0.8, 0.0}

\definecolor{americanrose}{rgb}{1.0, 0.01, 0.24}
\definecolor{cobalt}{rgb}{0.0, 0.28, 0.67}
\definecolor{brandeisblue}{rgb}{0.0, 0.44, 1.0}
\definecolor{mycolor}{rgb}{0.0, 0.0, 0.5}
\definecolor{oxfordblue}{rgb}{0.0, 0.13, 0.28}
\definecolor{azure}{rgb}{0.0, 0.5, 1.0}
\definecolor{turquoiseblue}{rgb}{0.0, 1.0, 0.94}
\newtcolorbox{mynewbox}[1]{colback=white!5!white,colframe=azure!75!black,fonttitle=\bfseries,title=#1}
\newtcolorbox{mybox}{colback=mycolor!5!white,colframe=azure!75!black}
\newtcolorbox{mynamedbox}[1]{colback=mycolor!5!white,colframe=azure!75!black,title=#1}
\definecolor{venetianred}{rgb}{0.78, 0.03, 0.08}
\newtcolorbox{mynamedbox1}[1]{colback=venetianred!5!white,colframe=venetianred!80!black,title=#1}
\newtcolorbox{mynamedbox2}[1]{colback=azure!5!white,colframe=azure!80!black,title=#1}

\definecolor{verdes}{rgb}{0.1, 0.5, 0.1}%
\definecolor{cornellred}{rgb}{0.7, 0.11, 0.11}

\definecolor{VioletRed4}{rgb}{0.55, 0.13, .32}

\hypersetup{
     colorlinks   = true,
     citecolor    = violet,
     urlcolor     = violet,
     linkcolor    = violet}

\definecolor{rossocorsa}{rgb}{0.83, 0.0, 0.0}

\def\lsim{\mathrel{\rlap{\lower4pt\hbox{\hskip0.5pt$\sim$}}
    \raise1pt\hbox{$<$}}}         
\def\gsim{\mathrel{\rlap{\lower4pt\hbox{\hskip0.5pt$\sim$}}
    \raise1pt\hbox{$>$}}}         

\usepackage[normalem]{ulem}

\begin{document}

\title[]{Nonlinear Effects in Black Hole Ringdown Made Simple:\\
Quasi-Normal Modes as Adiabatic Modes}

\author{A. Kehagias
\orcidlink{0000-0002-9552-9366}
}
\affiliation{Physics Division, National Technical University of Athens, Athens, 15780, Greece}
\affiliation{Department of Theoretical Physics and Gravitational Wave Science Center,  \\
24 quai E. Ansermet, CH-1211 Geneva 4, Switzerland}

\author{A. Riotto
\orcidlink{0000-0001-6948-0856}
}
\affiliation{Department of Theoretical Physics and Gravitational Wave Science Center,  \\
24 quai E. Ansermet, CH-1211 Geneva 4, Switzerland}


\begin{abstract}
\noindent
The nonlinear nature of general relativity manifests prominently throughout the merger of two black holes, from the inspiral phase to the final ringdown. Notably, the quasi-normal modes  generated during the ringdown phase display significant nonlinearities. We show that these nonlinear effects can be effectively captured by zooming in  on  the photon ring through the Penrose limit. Specifically, we model the quasi-normal modes as null particles trapped in unstable circular orbits around the black holes and show that they can be interpreted as adiabatic modes, perturbations that are arbitrarily close to  large diffeomorphisms. This enables the derivation of  a simple analytical expression for the QNM nonlinearities 
for Schwarzschild and Kerr black holes which   reproduces  well the existing numerical results.
\end{abstract}
\maketitle


\noindent
{\it Introduction.} The detection of quasi-normal modes (QNMs) from black hole ringdowns will enable consistency tests and independent verification of general relativity. So far, numerous studies have been conducted using current data \cite{LIGOScientific:2020tif,LIGOScientific:2021sio,Capano:2021etf, Finch:2022ynt, Isi:2022mhy, Cotesta:2022pci, Siegel:2023lxl} and making predictions for future ground- and space-based GW detectors (see, for instance, Refs.  \cite{Berti:2005ys, Ota:2019bzl, Bhagwat:2019dtm, Pitte:2024zbi}) based on linear ringdown QNM frequencies. However, given that general relativity is a nonlinear theory, recent investigations have verified the presence of quadratic QNMs in relativistic numerical simulations of binary black holes \cite{London:2014cma,Mitman:2022qdl, Cheung:2022rbm, Ma:2022wpv, Redondo-Yuste:2023seq, Cheung:2023vki, Zhu:2024rej}. Regarding their amplitudes, effort has aimed to derive the quadratic-to-linear amplitude ratio through both numerical and analytical methods \cite{Kehagias:2023ctr,Redondo-Yuste:2023seq, Cheung:2023vki,Perrone:2023jzq,Zhu:2024rej, Ma:2024qcv,Bourg:2024jme,Bucciotti:2024zyp,Khera:2024yrk}. For example, for the fundamental modes $(\ell=2, |m|=2, n=0)$ in a non-precessing and fully parity-even binary, the amplitude of the nonlinear sourced $\ell=4$ QNM, built up from two $\ell=2$ QNMs,  turns out to be for Schwarzschild black holes \cite{Redondo-Yuste:2023seq,Bucciotti:2024zyp}

\begin{eqnarray}\label{Qamp}
    {\cal R}_{2\times 2}=\left|\frac{A_{2\times 2}}{A_{2}^2}\right|\simeq 0.154.
\end{eqnarray}
This ratio is only mildly dependent upon the spin of the black hole and for highly spinning Kerr black holes it is reduced only by a factor $\sim 0.5$ \cite{Redondo-Yuste:2023seq}.
Such  nonlinear effect might  be observable in a few events using current ground-based detectors, with even more promising detection prospects for the LISA mission \cite{Yi:2024elj, Lagos:2024ekd}.

In this Letter we offer a simple and intuitive way to evaluate the nonlinearities of the QNMs. The key is to think of them as degrees of freedom which populate the photon ring and assume that the interactions sourcing the nonlinearities happen there. This amounts to studying the QNM dynamics not in the full original spacetime, but in the much simpler   plane wave spacetime resulting when we zoom in towards the photon ring through the so-called Penrose limit. This will allow to regard the QNMs as adiabatic modes, i.e. as  perturbations that are locally indistinguishable from a (large) change of coordinates and  to extract the nonlinear through a simple coordinate transformation.

In following  this simple line of thought, we will be able to provide a simple expression for the nonlinearity parameter ${\cal R}_{2\times 2}$, which reproduces remarkably well the numerical results 
for both Schwarzschild and Kerr black holes.


\vskip 0.5cm
\noindent
{\it The general argument.}
The Penrose limit \cite{Reidel:1976} associates to every space-time metric $g_{ab}$, with line element ${\rm d}s^2$,  and a  null geodesic $\gamma$ in that space-time a (limiting) plane wave metric. The first step is to rewrite the metric in coordinates ``adapted" to $\gamma$, which corresponds to an embedding of $\gamma$ into a twist-free congruence of null geodesics, given by some coordinates $V$ and $Z$ and $\bar Z$ constant, with the remaining coordinate $U$ playing the role of the affine parameter and $\gamma(U)$ coinciding with the geodesic at $V=Z=\bar Z=0$. The next step is to perform the change of coordinates $(U,V,Z,\bar Z)=(u,\lambda^2 v,\lambda z,\lambda \bar z)$ for some real $\lambda$ and to take the  limit
    $\lim_{\lambda\rightarrow 0}\lambda^{-2}{\rm d} s^2={\rm d}s^2_\gamma$. Notice that the operation of rescaling the metric is priceless as we are zooming in  on null geodesics.

The resulting metric is the so-called
  Penrose limit of the initial spacetime which,  recasted in Brinkmann coordinates, has the pp-wave form

\begin{equation}
\label{metric}
{\rm d} s^2_\gamma
=2{\rm d}u{\rm d}v+H(u,z,\bar z) {\rm d}u^2-2{\rm d}z{\rm d}\bar z.
\end{equation}
The coordinate $u$ plays the role of the affine parameter along the geodesic and the function $H$ controls the  geodesic deviation properties along the transverse coordinates $z$ and $\bar z$.

Our argument hinges on three key elements. We start from the realization that the QNMs  of black holes  are determined by the parameters of the circular null unstable geodesics. These ``free" modes of vibration can be interpreted in terms of null particles trapped at the unstable circular orbit and slowly leaking out (see,  for instance, Ref. \cite{Cardoso:2008bp}). We will assume therefore that the nonlinearities of the QNMs are mainly generated at the photon ring.

 The second key ingredient  is that, for a generic pp-wave spacetime, like the one described by the metric (\ref{metric}) and obeying the condition (from now on subscripts indicate derivatives with respect to the indicated variable, e.g. $H_z=\partial_z H$ and so on)

\begin{equation}
 H_{z\bar z}=0,
\end{equation}
which is true for Schwarzschild and Kerr black holes, 
any metric perturbation $h_{ab}=(\gamma_{ab}+\bar\gamma_{ab}) $ satisfying the linearized Einstein vacuum equations may be described   by a  complex scalar field $\Phi$, such that in the radiation gauge $\ell^a h_{ab}=g^{ab}h_{ab}=0$ (the superscript ${}^{(0)}$ indicates the zero-th order) 
\cite{Araneda:2022lgu}

\begin{equation}
\gamma_{ab}= \Phi_{\bar z\bar z} \ell_a^{(0)}\ell_b^{(0)}+2 \Phi_{v\bar z}\ell^{(0)}_{(a}m^{(0)}_{b)}+ \Phi_{vv} m^{(0)}_a m^{(0)}_b.
\end{equation}
The complex null tetrad $(\ell,n,m,\bar m)$ is defined in the Supplementary Material (SM) at zero-th and first-order.
The background plus linearly perturbed metric  acquires  the form    

\begin{eqnarray}
\label{metricp}
{\rm d} s^2&=&2{\rm d}u{\rm d}v+\left[H+\Phi_{\bar z\bar z}+\bar \Phi_{zz}\right] {\rm d}u^2\nonumber\\
&+&
\Phi_{v\bar z}{\rm d} u{\rm d}z+\bar \Phi_{v z}{\rm d} u{\rm d}\bar z
\nonumber\\
&+& \Phi_{vv}{\rm d} z^2+
\bar \Phi_{vv}{\rm d} \bar z^2
-2{\rm d}z{\rm d}\bar z.
\end{eqnarray}

The scalar field satisfies the wave equation

\begin{equation}
\label{equation}
\Box \Phi=2(\Phi_{uv} -\Phi_{z\bar z})-H\Phi_{vv}=0,
\end{equation} 
whose solution, by properly choosing the boundary conditions,  describes a QNM, 

\begin{eqnarray}
\label{nonlimit}
    \Phi(u,v,z,\bar z)&=&A e^{i P_u u+i P_v v}\chi(z,\bar z),\nonumber\\
    \ln \chi(z,\bar z)&=&\alpha_{zz} z^2+\alpha_{\bar z\bar z} \bar z^2+\alpha_{z\bar z} z\bar z.
\end{eqnarray}
The third ingredient is that for any solution $\Phi(u,v,z,\bar z)$ of such wave equation, also $\Phi(u,\lambda^2 v,\lambda z,\lambda\bar z)$ is a solution
as long as $H(u,\lambda z,\lambda\bar z)=\lambda^2H(u, z,\bar z)$ (again true for Schwarzschild and Kerr black holes). 

By taking the limit of small $\lambda$, we  can therefore further zoom in  towards    the photon ring. In doing so the linear perturbation becomes only a function of the affine parameter,    

\begin{eqnarray}
\label{limit}
    \Phi(u,v,z,\bar z)&\to&\phi(u)\equiv A e^{i P_u u},\nonumber\\
    \Phi_{\bar z\bar z}(u,v,z,\bar z)&\to &2\alpha_{\bar z\bar z}\phi(u),\nonumber\\
    \Phi_{ z z}(u,v,z,\bar z)&\to &2\alpha_{zz}\phi(u),\nonumber\\
\Phi_{v z}(u,v,z,\bar z)=&\to &0,\nonumber\\
\Phi_{vv}(u,v,z,\bar z)&\to&-P_v^2\phi(u)
\end{eqnarray}
and it represents a wave propagating
along the photon ring with a given frequency (see the SM for a more rigorous demonstration of this statement). 

Thanks to the residual gauge transformations in the radiation gauge (see the SM details), we can remove such a wave by a convenient set of coordinate transformations 

\begin{eqnarray}
\label{transformation}
  z&\to &z(1+P_v^2 \phi_u),\nonumber\\
  v&\to &v+\frac{P_v^2}{2} z^2\phi_u+g(u),\nonumber\\
  g(u)&=&-\frac{\alpha_{\bar z\bar z}}{2}\int^u{\rm d}u' \phi(u')+{\rm c.c.}
\end{eqnarray}
and take the limit of vanishing $z$. This means that, zooming in towards the photon ring, 
the linear perturbations  become arbitrarily close to some large gauge transformation (a diffeomorphism that does not vanish at spatial infinity).

Readers with some prior experience in  cosmology will recognize the similarity with the  ``adiabatic" modes introduced by 
Weinberg \cite{Weinberg:2003sw}, that is  physical, finite momentum perturbations around an FLRW background that, in the zero momentum limit, become arbitrarily close to some large gauge transformation. The operation of taking the  long wavelength limit of the cosmological perturbations is equivalent  to zooming in towards the photon ring.  The perturbation (\ref{limit}) is an adiabatic mode as it can be promoted to to a non-trivial perturbation (\ref{nonlimit}) away from the photon ring.

Imagine now that there is a fundamental QNM with multipole $\ell=2$ living on the photon ring. Zooming in on the latter and changing properly the coordinates we can go back to the black hole background metric (\ref{metric}) in these new coordinates. On such a background, we insert again an $\ell=2$ QNM,  whose profile will be again of the form (\ref{nonlimit}), but this time in the new coordinates. Expanding back in the old coordinates we will generate a quadratic $2\times 2$ mode, that is  a nonlinear  $\ell=4$ QNM from the nonlinearities of gravity.

We now proceed with analyzing the Schwarzschild and Kerr black hole, where  the procedure we have briefly described will become hopefully more transparent.
\vskip 0.5cm
\noindent
{\it The Schwarzschild black hole.} 
The Schwarzschild black hole of mass $M$ induces a metric

\begin{eqnarray}
  {\rm d}s^2&=&-f(r) {\rm d}t^2+f^{-1}(r){\rm d}r^2+r^2({\rm d}\theta^2+\sin^2\theta{\rm d}\phi^2),\nonumber\\
  f(r)&=&1-2M/r,
\end{eqnarray}
where we have set the Newton constant to unity.
The Penrose limit  around the circular photon ring located at $r_0=3M$ and $\theta=\pi/2$ has the metric (\ref{metric}) with

\be
H(z,\bar z)=\frac{1}{3M^2}(z^2+\bar z^2).
\ee 
For the QNMs, we first look for linear solutions to the Eq. (\ref{equation}) and  demand an outgoing boundary condition for the unstable direction $\delta r=r-r_0=(z+\bar z)/\sqrt{6}$, and, additionally, we impose a decaying boundary condition for the stable direction $\delta\theta=\theta-\pi/2=i(\bar z-z)/r_0\sqrt{2}$. Combined, these yield the quantization condition

\begin{equation}
    P_u=\frac{i}{2\sqrt{3}M}-\frac{1}{2\sqrt{3}M}.
\end{equation}
On the other hand, periodicity along the azimuthal angle imposes

\begin{equation}
    P_v=\ell\omega_{r}=\frac{\ell}{3\sqrt{3}M},
\end{equation}
where $\omega_r$ is the orbital frequency, $\ell=m$ is the angular momentum,  and for the 
Schwarzschild black hole is equal to the precession frequency $\omega_p$ around the original orbit.

The linear fundamental  QNMs have  the following solution \cite{Fransen:2023eqj,Giataganas:2024hil}

\begin{eqnarray}
    \Phi_\ell(u,v,z,\bar z)&=&A_\ell e^{i P_u u+i P_v v}\chi_\ell(z,\bar z),\nonumber\\
    \ln\chi_\ell(z,\bar z)&=&\frac{3}{4}\ell\omega_r\left[
    (\omega_p+i\lambda_{\text{\tiny  L}})(z^2+\bar z^2)\right.\nonumber\\
    &+&\left. 2i(\lambda_{\text{\tiny  L}}+i\omega_p)z\bar z\right],
\end{eqnarray}
where $\lambda_{\text{\tiny  L}}=\omega_r$ is the characteristic Lyapunov timescale.

Consider now a QNM with multipole $\ell$ living on the photon ring and imagine to zoom in towards the photon ring as explained in the previous section. By the residual change of coordinates (\ref{transformation})
we can eliminate such linear QNM and go back
to the original background metric. In particular, the coordinate $v$ transforms as $v\to v'=v+g(u)$, with

\begin{equation}
\label{gu}
g(u)
=-i\frac{ \alpha_{\bar z\bar z}}{P_u} A_\ell e^{iP_u u} +{\rm c.c.}  
\end{equation}
In this new background, close 
 to the photon ring, we insert another QNM with multipole $\ell$, whose   profile will be

\begin{eqnarray}
\Phi_\ell(u,v')&\simeq& A_\ell\exp\left(i P_u u +iP_v v'\right)\nonumber\\
&=&A_\ell\exp\big(i P_u u+ iP_v v
\nonumber\\
&+&A_\ell\frac{P_v}{P_u} \alpha_{\bar z\bar z}
e^{iP_u u} +{\rm c.c.}\big)
\end{eqnarray}
and therefore  implicitly depending upon the original  QNM whose presence is now encoded in the new coordinate system thanks to the residual gauge symmetry. 

 Expanding up to  second-order in the linear amplitude $A_\ell$ delivers a propagating   $2\ell$-mode   built up from two $\ell$-modes
 in the original coordinate system

\begin{equation}
\Phi_{\ell\times \ell}\simeq A_\ell^2\frac{ P_v}
{P_u} \alpha_{\bar z\bar z} e^{2iP_u u}+{\rm constant},
\end{equation}
where the unphysical constant is  eliminated when  requiring that the second-order perturbation must have  vanishing expectation value. Recalling that the two gravitational polarizations $h_{+,\times}$ are proportional to the real
and imaginary parts of the complex scalar  field $\Phi$, we get that 
the ratio ${\cal R}_{\ell \times \ell}$ for them   is 
\begin{eqnarray}
{\cal R}_{\ell \times \ell}=2\left|\frac{P_v\alpha_{\bar z\bar z}}{P_u}\right|,
\label{RR}
\end{eqnarray}
where

\begin{eqnarray}
\alpha_{\bar z\bar z}&=&\frac{3}{4} \ell \omega_r (\omega_p+i\lambda_{\text{\tiny  L}}), \quad 
P_u=\frac{3}{2}(i\lambda_{\text{\tiny  L}}-\omega_p), \nonumber\\
P_v&=&\omega_r \ell.
\end{eqnarray}
For the Schwarzschild black hole we have 
\begin{eqnarray}
\omega_r=\omega_p=\lambda_{\text{\tiny  L}}=\frac{1}{3\sqrt{3}M}
\end{eqnarray}
and the  ratio  (\ref{RR}) turns out to be 
\begin{eqnarray}
\label{final}
{\cal R}_{\ell \times \ell}=\ell^2 \omega_r^2.
\end{eqnarray}
For $\ell=m=2$ we finally find (in units of $M=1$)

\begin{eqnarray}
{\cal R}_{2\times 2}=\frac{4}{27}\simeq 0.15.
\end{eqnarray}
This result aligns remarkably well  with the value reported in Ref. \cite{Redondo-Yuste:2023seq,Bucciotti:2024zyp} for spinless black holes. This might sound  surprising, considering that the Penrose limit is primarily intended to describe the eikonal regime of  the QNM properties. However, it is important to note that even for the $\ell=2$ mode, this approximation yields the fundamental frequency with an error margin of only  ${\cal O}(20\%)$.

\vskip 0.5cm
\noindent
{\it The Kerr black hole.} For the Kerr black holes, our logic goes through as for the
Schwarzschild black holes and the final formula (\ref{RR}) may be  applied as well.
The expressions fo the frequencies and the Laypunov timescale   are  rather unwieldy because 
the general plane wave metric (\ref{metric}) for Kerr black holes is time-dependent and not diagonal. The time-dependence comes from the dependence $\theta = \theta(u)$. However, for the equatorial orbits, the problem simplifies and \cite{Fransen:2023eqj}

\begin{eqnarray}
    \omega_r(a)&=&\frac{3M-r_0(a)}{a[r_0+3M(a)]},\nonumber\\
\omega^2_p(a)&=&\lambda^2_{\text{\tiny  L}}(a)=\frac{12 M\Delta(a)}{r_0^3(a)[r_0+3M(a)]^2},\nonumber\\
\Delta(a)&=&r_0^2(a)-2M r_0(a)+a^2,
\end{eqnarray}
 where $a$ is the spin of the black hole and 
$r_0(a)$ satisfies the equation 

\begin{equation}
    r_0(r_0-3M)^2-4 a^2 M=0.
\end{equation}
\begin{figure}[t]
    \centering
    \includegraphics[width=0.5\textwidth]{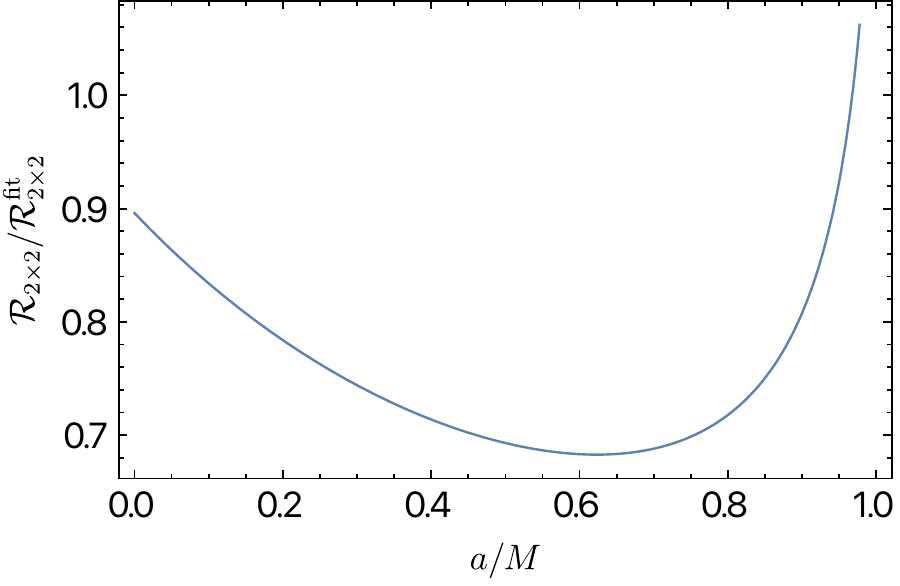}
    \caption{ratio between our result
    (\ref{finalKerr}) and the fit provided in Ref. \cite{Redondo-Yuste:2023seq}.}
    \label{fig:1}
\end{figure}
\noindent
 For  $\ell=m=2$ the  ratio  (\ref{RR}) turns out to be  (again in units of $M=1$)
\begin{eqnarray}
\label{finalKerr}
{\cal R}_{2 \times 2}=4 \omega_r^2(a).
\end{eqnarray}
For co-rotating orbits (the counter-rotating ones are  increasingly suppressed for large values of $a/M$) one has 

\begin{equation}
\label{ratio}
    \frac{\omega_r(a/M\simeq 1) }{\omega_r(a/M\simeq 0)}\simeq \frac{3}{4}.
\end{equation}
 Since the  orbital frequency is a function that mildly depends upon the spin of the black hole, so does the ratio (\ref{finalKerr}), which  decreases only by a factor $\sim (3/4)^2\sim 0.5$,  as observed in Ref.  \cite{Redondo-Yuste:2023seq}. In the same reference the authors provided a fit to the quantity ${\cal R}_{2 \times 2}$ obtained using second-order in black hole perturbation theory, 
 
 \begin{eqnarray}
     {\cal R}_{2 \times 2}^{\text{\tiny  fit}}&=&0.058+[\kappa(a)M]^{1.61},\nonumber\\
     \kappa(a)&=&\sqrt{M^2-a^2}/[2M(M+\sqrt{M^2-a^2})].
 \end{eqnarray}
 We plot in Fig. \ref{fig:1} the ratio between our result and such a fit. It is quite satisfying that our simple approach reproduces the fit with, at worst, an accuracy of ${\cal O}(30\%)$, especially considering that we only included contributions from equatorial orbits.

\vskip 0.5cm
\noindent
{\it Conclusions.} We have shown that the nonlinear behavior of black hole QNMs can be effectively understood through the Penrose limit, with a particular focus on adiabatic modes. By zooming in on the photon ring, we revealed that linear perturbations  approach adiabatic modes and become arbitrarily close to large gauge transformations. This insight is critical, as it allows us to understand how the gravitational nonlinearities manifest through the generation of higher-order QNMs from the fundamental mode. Furthermore, we applied this framework to the Schwarzschild and Kerr black holes, confirming that the nonlinear mode amplitudes fit well with previous results.  
The key role played by adiabatic modes emphasizes the importance of treating black hole perturbations not just as linear excitations but as dynamic systems that, when viewed in the proper limit, reveal deeper connections to the spacetime's structure and its nonlinear nature. This understanding lays a crucial groundwork for future investigations into gravitational wave signatures, where these nonlinear, adiabatic modes may become observable in black hole mergers.

There is potential to improve  the approach proposed in this Letter. Our calculation has been restricted to reproducing the 
for the fundamental modes $(\ell=2, |m|=2, n=0)$  in
a non-precessing and fully parity-even binary, relying on the assumption of equatorial symmetry for the QNM amplitudes. If we relax this assumption, each individual $(\ell, m)$ linear metric perturbation will include both even and odd components. Consequently, the quadratic modes cannot be fully described by the single parameter ${\cal R}_{2\times 2}$, although deviations are limited to be  ${\cal O}(20\%)$ \cite{Bourg:2024jme,Bucciotti:2024jrv}. Extending our method to account for the complete spectrum of nonlinearities, zooming in onto different QNM multipoles,  would be a valuable next step. It would also be intriguing to further explore the connection between zoomed-in QNMs and the adiabatic modes by investigating whether consistency relations, akin to those found in cosmological contexts \cite{Maldacena:2002vr,Kehagias:2013yd,Peloso:2013zw}, can be derived among the various nonlinearities.

\begin{acknowledgments}
\vspace{5pt}\noindent\emph{Acknowledgments --}
We than E. Berti, V. Cardoso and G. Carullo for useful discussions and feedback on the draft.  A.R.  acknowledges support from the  Swiss National Science Foundation (project number CRSII5\_213497) and from   the Boninchi Foundation for the project ``PBHs in the Era of GW Astronomy''.  A.K. acknowledges support from the Swiss National
Science Foundation (project number IZSEZ0 229414).

\end{acknowledgments}

\bibliography{Draft}

\clearpage
\newpage
\maketitle
\onecolumngrid
\begin{center}
\vspace{0.05in}
 
\vspace{0.05in}
{ \large\it Supplementary Material}
\end{center}
\onecolumngrid
\setcounter{equation}{0}
\setcounter{figure}{0}
\setcounter{section}{0}
\setcounter{table}{0}
\setcounter{page}{1}
\makeatletter
\renewcommand{\theequation}{S\arabic{equation}}
\renewcommand{\thefigure}{S\arabic{figure}}
\renewcommand{\thetable}{S\arabic{table}}
\subsection{The  complex null tetrad up to linear order}
\noindent
The first-order perturbed metric in Eq. (\ref{metricp}) can be written as 
\begin{eqnarray}
    g_{ab}=2
    \ell_{(a}n_{b)}-2 m_{(a}\bar m_{b)}, 
\end{eqnarray}
where 
\begin{eqnarray}
    \ell_a=\ell^{(0)}_a+\ell^{(1)}_a,
    \qquad n_a=n^{(0)}_a+n^{(1)}_a, \qquad
    m_a=m^{(0)}_a+m^{(1)}_a, \qquad
    \bar m_a=\bar m^{(0)}_a+\bar m^{(1)}_a,
\end{eqnarray}
Here 
\begin{eqnarray}
\ell^{(0)}_a {\rm d}x^a &=& {\rm d}u, \,\, n^{(0)}_a {\rm d}x^a = H{\rm d}u + {\rm d}v, 
m^{(0)}_a {\rm d}x^a = {\rm d}z, \,\, \bar{m}^{(0)}_a {\rm d}x^a = {\rm d}\bar{z}.
\end{eqnarray}
so that

\begin{eqnarray}
\ell_{(0)}^a\partial_a&=&\partial_v,\,\,\,
m_{(0)}^a\partial_a=-\partial_{\bar z},\,\,\,
\bar m_{(0)}^a\partial_a=-\partial_{z},\,\,\,
n_{(0)}^a\partial_a=\partial_{u}-\frac{H}{2}\partial_v,
    \end{eqnarray}
At first-order we have

\begin{align}
    \ell_a^{(1)}=&(0,0,0,0),\,\,\,
    n^{(1)}_a=\frac{1}{2}(\Phi_{\bar z\bar z}+\bar \Phi_{zz})\, \ell_a^{(0)}, \nonumber\\
    m^{(1)}_a=&-\frac{1}{2}\Phi_{v\bar z}\ell_a^{(0)}-\frac{1}{2}\Phi_{vv}\, \bar m_a^{(0)},\,\,\,
    \bar m^{(1)}_a=-\frac{1}{2}\bar\Phi_{v  z}\ell_a^{(0)}-\frac{1}{2}\bar \Phi_{vv}\,  m_a^{(0)} .
\end{align}
Then it turns out that the Weyl scalars are
\begin{align}
\Psi_0^{(0)}=&\Psi_1^{(0)}=\Psi_2^{(0)}=\Psi_3^{(0)}=0,\,\,\,
\Psi_4^{(0)}=\frac{1}{2}H_{zz},
    \end{align}
and 
\begin{align}
    \Psi_0^{(1)}=&\frac{1}{2}\bar \Phi_{vvvv}, \nonumber\\
    \Psi_1^{(1)}=& -\frac{3}{8}\bar \Phi_{vvvz}, \nonumber \\
    \Psi_2^{(2)}=&\frac{1}{12}\Phi_{vv\bar z\bar z}+\frac{1}{3}\bar \Phi_{vvzz}, \label{psi1} \\
    \Psi_3^{(1)}=& -\frac{1}{8}  \Phi_{vz\bar z\bar z}-\frac{3}{8} \bar \Phi_{vzzz}+\frac{1}{16} H\Phi_{vvv\bar z}-\frac{1}{8}\Phi_{uvv\bar z}\nonumber\\
    \Psi_4^{(1)}=&\frac{1}{2}\Phi_{zz\bar z \bar z}+\frac{1}{2} \bar \Phi_{zzzz}+\frac{1}{4}
    H_{\bar z} \Phi_{vvz}+\frac{1}{4} H \Phi_{vvz\bar z}
    -\frac{1}{4}H_u \Phi_{vvv}+\frac{1}{8}H^2 \Phi_{vvvv}-\frac{1}{2}\Phi_{uvz\bar z}-\frac{1}{2} H \Phi_{uvvv}+\frac{1}{2}\Phi_{uuvv}\nonumber .
\end{align}

    \subsection{Linear gauge transformation}
\noindent
Changing the coordinates at the linear level as $x_a\rightarrow x_a+\xi_a$, For an arbitrary covector field $\xi_{a}$, we define the ``gauge 
operator'' (or Killing operator) $K$ by
\begin{equation}
 K[\xi]_{ab} = \nabla_{a}\xi_{b} + \nabla_{b}\xi_{a}.
 \label{GaugeOperator}
\end{equation}
In terms of a null tetrad, this can be written as follows
(we drop the labels to indicate the order of the perturbation) \cite{Araneda:2022lgu}

\begin{align}
 K[\xi]_{ab} ={}& K_{nn}\ell_{a}\ell_{b} -2K_{n\bar{m}}\ell_{(a}m_{b)}
-2K_{nm}\ell_{(a}\bar{m}_{b)} 
+K_{\bar{m}\bar{m}}m_{a}m_{b} 
+K_{mm}\bar{m}_{a}\bar{m}_{b} \\
 & + K_{\ell\ell} n_{a}n_{b} -2K_{\ell\bar{m}}n_{(a}m_{b)} 
 -2K_{\ell m}n_{(a}\bar{m}_{b)} +2K_{\ell n} n_{(a}\ell_{b)} 
-2K_{m\bar{m}}m_{(a}\bar{m}_{b)}.
\end{align}
For a pp-wave, we find (we return to the usual notation of the derivatives to avoid cluttering notation)

\begin{subequations}\label{GOcomponents}
\begin{align}
 & K_{\ell\ell} = 2\partial_{v}\xi_{\ell}, 
 &
 &K_{m\bar{m}} =
 -(\partial_{\bar{z}}\xi_{\bar{m}}+\partial_{z}\xi_{m})
 \\
 &K_{\ell\bar{m}} = \partial_{v}\xi_{\bar{m}} 
 - \partial_{z}\xi_{\ell}, 
 &
 & K_{nm} =  (\partial_{u}-\tfrac{1}{2}H\partial_{v})\xi_{m} 
 +\bar{\kappa}'\xi_{\ell} - \partial_{\bar{z}}\xi_{n}, 
 \\
 &K_{\ell m} = \partial_{v}\xi_{m} - \partial_{\bar{z}}\xi_{\ell}, 
 &
 & K_{n\bar{m}} =(\partial_{u}-\tfrac{1}{2}H\partial_{v})\xi_{\bar{m}} 
 +\kappa'\xi_{\ell} - \partial_{z}\xi_{n}, 
 \\
 & K_{\bar{m}\bar{m}} = -2\partial_{z}\xi_{\bar{m}}, 
 &
 & K_{\ell n} = 
 \partial_{v}\xi_{n} + (\partial_{u}-\tfrac{1}{2}H\partial_{v})\xi_{\ell}, 
 \\
 & K_{mm} = -2\partial_{\bar{z}}\xi_{m}, 
 &
 & K_{nn} = 2(\partial_{u}-\tfrac{1}{2}H\partial_{v})\xi_{n} 
 +2\kappa'\xi_{m} + 2\bar{\kappa}'\xi_{\bar{m}},
\end{align}
\end{subequations}
with 

\begin{equation}
    \kappa'=\frac{1}{2}\partial_z H.
\end{equation}
\vskip 0.2cm
\centerline{\it Residual gauge transformations}
\vskip 0.5cm
\noindent
The radiation gauge  is preserved by 
transformations in which the new gauge vector $\xi_{a}$ satisfies
\begin{equation}
 \ell^{a}K[\xi]_{ab} = 0, \qquad g^{ab}K[\xi]_{ab} = 0,
 \label{RGresidual}
\end{equation}
or 
equivalently

\begin{equation}
K_{\ell\ell}=K_{\ell m}=K_{\ell \bar{m}}=K_{\ell n} 
=K_{m\bar{m}}=0,
\end{equation}
where $K_{\ell\ell}=\ell^a\ell^aK_{ab}$, $\xi_m=m^a\xi_a$, and so on.
Using the identities \eqref{GOcomponents}, this is
\begin{subequations}\label{RGresidualComponents}
\begin{align}
 \partial_{v}\xi_{\ell} ={}& 0, \\
 \partial_{v}\xi_{\bar{m}} - \partial_{z}\xi_{\ell} ={}& 0, \\
 \partial_{v}\xi_{m} - \partial_{\bar{z}}\xi_{\ell} ={}& 0, \\
 \partial_{v}\xi_{n} + \partial_{u}\xi_{\ell} ={}& 0, \\
 \partial_{\bar{z}}\xi_{\bar{m}} + \partial_{z}\xi_{m} ={}& 0.
\end{align}
\end{subequations}
From here we deduce 
\begin{equation}
 \partial^{2}_{v}\xi_{\bar{m}} = \partial^{2}_{v}\xi_{m} 
 = \partial^{2}_{v}\xi_{n} = 0, \qquad 
 \partial_{z}\partial_{\bar{z}}\xi_{\ell} = 0. 
 \label{SecondDerivativesGV}
\end{equation}
Notice that in view of \eqref{GOcomponents}, 
and given that $\partial_{v}$ is a Killing vector of the background 
space-time, from \eqref{SecondDerivativesGV} it follows that
$\partial^{2}_{v}K_{ ab}=0$ for all $( a,b)=(u,v,z,\bar{z})$. 
The components of the metric under the residual transformations transform as

\begin{eqnarray}
    \Phi_{\bar z\bar z}&\rightarrow& \Phi_{\bar z\bar z}+(\partial_{u}-\tfrac{1}{2}H\partial_{v})\xi_{n}+\kappa'\xi_{m},\nonumber\\
    \Phi_{v\bar z}&\rightarrow&\Phi_{v\bar z}-(\partial_{u}-\tfrac{1}{2}H\partial_{v})\xi_{\bar{m}} 
 +\kappa'\xi_{\ell} + \partial_{z}\xi_{n},\nonumber\\
\Phi_{vv}&\rightarrow&\Phi_{vv}-\partial_{z}\xi_{\bar{m}}.
\label{Ss}
\end{eqnarray}
In the limit in which we zoom in on the photon ring resulting in the elements of the perturbed metric (\ref{limit}), the perturbations may be removed properly by choosing

\begin{eqnarray}
  \xi_{\bar{m}}&=&P_v^2 z\phi(u),\nonumber\\
  \xi_n&=&\frac{P_v^2}{2} z^2\partial_u\phi(u)+\chi(u),\nonumber\\
  \chi(u)&=&-\frac{\alpha_{\bar z\bar z}}{2}\int{\rm d}u' \phi(u')
\end{eqnarray}
and taking the limit of vanishing $z$.

\subsection{QNMs propagating along the photon ring}
\noindent
We  apply the Penrose limit to  the metric (\ref{metricp}), which is already and conveniently expressed in ``adapted" coordinates. The 
null coordinate vector field $\partial_v$ is geodesic and it is in this sense that the coordinates are adapted to the null geodesic congruence generated by $\partial_v$. Explicitly, the Penrose limit of the metric with respect to the null geodesic  with affine parameter $v$ at $u=z=\bar z=0$ yields the metric

\begin{eqnarray}
\label{metricpPL}
{\rm d} s^2&=&2{\rm d}u{\rm d}v-2{\rm d}z{\rm d}\bar z\nonumber\\
&+& \phi_{vv}(v){\rm d} z^2+
\bar \phi_{vv}(v){\rm d} \bar z^2\nonumber\\
&\equiv & 2{\rm d}u{\rm d}v+g_{ij}(v){\rm d} z^i {\rm d} z^j,
\end{eqnarray}
where $z^i=(z,\bar z)$ and 

\begin{equation}
    \phi_{vv}(v)=\left.\Phi_{vv}(u,v,z,\bar z)\right|_{u=z=\bar z=0}.
    \end{equation}
The last step is to recast the metric (\ref{metricpPL}) in Brinkmann coordinates. We set

\begin{eqnarray}
 z^i&=&F^i_a(v)Z^a,
\end{eqnarray}
where  $Z^a=(Z,\bar Z)$, such that

\begin{equation}
g_{ij}F^i_a F^j_b{\rm d} Z^a{\rm d} Z^b=-2{\rm d} Z{\rm d} \bar Z.
\end{equation}
Since

\begin{eqnarray}
 {\rm  d}z^i&=&F^i_a(v){\rm  d}Z^a+ 
 F^i_{a,v}(v)Z^a{\rm d} v,
 \end{eqnarray}
we have

\begin{eqnarray}
    g_{ij} {\rm  d}z^i{\rm  d}z^j&=&g_{ij}\left(F^i_a{\rm  d}Z^a+ 
 F^i_{a,v}Z^a{\rm d} v\right)\left(F^j_b{\rm  d}Z^b+ 
 F^j_{b,v}Z^b{\rm d} v\right)\nonumber\\
 &=&-2{\rm d} Z{\rm d} \bar Z+g_{ij}F^i_aF^j_{b,v}Z^b{\rm  d}Z^a{\rm d} v
 +g_{ij}F^j_bF^i_{a,v}Z^a{\rm  d}Z^b{\rm d}v\nonumber\\
 &+&g_{ij}F^i_{a,v}F^j_{b,v}Z^a Z^b{\rm d}v^2.
\end{eqnarray}
Furthermore, we take

\begin{eqnarray}
   v&=&V,\nonumber\\
   u&=&U-\frac{1}{2}F_{ai,v}F^{i}_b Z^aZ^b=U-\frac{1}{2}g_{ij}F^j_{a,v}F^{i}_b Z^aZ^b, \nonumber\\
   &&
\end{eqnarray}
for which

\begin{eqnarray}
   2{\rm d} u{\rm d} v&=&2{\rm d}U{\rm d} V\nonumber\\
   &-&F_{ai,vv}F^{i}_b Z^aZ^b {\rm d}v^2-F_{ai,v}F^{i}_{b,v} Z^aZ^b {\rm d}v^2\nonumber\\
   &-&F_{ai,v}F^{i}_b Z^a{\rm d}Z^b{\rm d} v-F_{ai,v}F^{i}_b Z^b{\rm d}Z^a {\rm d} v. \nonumber\\
   &&
\end{eqnarray}
The resulting metric is therefore again of the pp-wave form

\begin{eqnarray}
\label{metricnew}
 {\rm d}s^2&=&2{\rm d}U{\rm d} V + 
 \mathscr{H}(V,Z,\bar Z){\rm d}V^2-2{\rm d}Z{\rm d}\bar Z,\nonumber\\
 &&
\end{eqnarray}
 with 

\begin{equation}
  \mathscr{H}(V,Z,\bar Z)=-F_{ai,VV}F^{i}_b Z^aZ^b=F^j_{a,VV}g_{ij}F^{i}_bZ^aZ^b.  
\end{equation}
At the linear order in $\phi(V)$ (and its conjugate) it is easy to demonstrate that

\begin{equation}
   F^i_a=\left(
   \begin{array}{cc}
1 & \bar \phi_{VV}(V)/2\\
 \phi_{VV}(V)/2 & 1
   \end{array}
   \right),
\end{equation}
so that

\begin{equation}
  \mathscr{H}(V,Z,\bar Z)=\frac{1}{2}\left[\phi_{VVVV}(V)Z^2+\bar \phi_{VVVV}(V)\bar Z^2\right].  
\end{equation}
This is the metric of a plane wave in Brinkmann coordinates and 
setting $\phi_{VVVV}=(h_++i h_\times)$, the wave profile  takes the form of the two gravitational wave polarizations.

\vskip.4in

\end{document}